\documentclass[authorversion,screen]{acmart}

\usepackage{times}
\usepackage{helvet}
\usepackage{courier}
\usepackage{graphicx}
\usepackage{comment}

\AtBeginDocument{%
  \providecommand\BibTeX{{%
    \normalfont B\kern-0.5em{\scshape i\kern-0.25em b}\kern-0.8em\TeX}}}
\setcopyright{none}
\copyrightyear{}
\acmYear{}
\begin{abstract}
\begin{quote}
Fairness is central to the ethical and responsible development and use of AI systems, with a large number of frameworks and formal notions of algorithmic fairness being available. 
However, many of the fairness solutions proposed revolve around technical considerations and not the needs of and consequences for the most impacted communities. 
We therefore want to take the focus away from definitions and allow for the inclusion of societal and relational aspects to represent how the effects of AI systems impact and are experienced by individuals and social groups.
In this paper, we do this by means of proposing the ACROCPoLis framework to represent allocation processes with a modeling emphasis on fairness aspects. 
The framework provides a shared vocabulary in which the factors relevant to fairness assessments for different situations and procedures are made explicit, as well as their interrelationships.
This enables us to compare analogous situations, to highlight the differences in dissimilar situations, and to capture differing interpretations of the same situation by different stakeholders.
 
\end{quote}
\end{abstract}



\begin{document}

\title{ACROCPoLis: A Descriptive Framework for Making Sense of Fairness}

\author{Andrea Aler Tubella}
\authornote{All authors contributed equally to this research. Authors listed alphabetically}
\email{andrea.aler@umu.se}
\affiliation{%
  \institution{Ume{\aa} University}
  \country{Sweden}
}

\author{Dimitri Coelho Mollo}
\email{dimitri.mollo@umu.se}
\affiliation{%
  \institution{Ume{\aa} University}
  \country{Sweden}
}

\author{Adam Dahlgren Lindstr\"{o}m}
\email{dali@cs.umu.se}
\affiliation{%
  \institution{Ume{\aa} University}
  \country{Sweden}
}

\author{Hannah Devinney}
\email{hannahd@cs.umu.se}
\affiliation{%
  \institution{Ume{\aa} University}
  \country{Sweden}
}

\author{Virginia Dignum}
\email{virginia.dignum@umu.se}
\affiliation{%
  \institution{Ume{\aa} University}
  \country{Sweden}
}

\author{Petter Ericson}
\email{pettter@cs.umu.se}
\affiliation{%
  \institution{Ume{\aa} University}
  \country{Sweden}
}

\author{Anna Jonsson}
\email{aj@cs.umu.se}
\affiliation{%
  \institution{Ume{\aa} University}
  \country{Sweden}
}

\author{Timotheus Kampik}
\email{tkampik@cs.umu.se}
\affiliation{%
  \institution{Ume{\aa} University \& SAP Signavio}
  \country{Sweden}
}

\author{Tom Lenaerts}
\email{Tom.Lenaerts@ulb.be}
\affiliation{%
  \institution{Universit{\'e} Libre de Bruxelles \& Vrije Universiteit Brussel}
  \country{Belgium}
}

\author{Julian Alfredo Mendez}
\email{julian.mendez@cs.umu.se}
\affiliation{%
  \institution{Ume{\aa} University}
  \country{Sweden}
}

\author{Juan Carlos Nieves}
\email{jcnieves@cs.umu.se}
\affiliation{%
  \institution{Ume{\aa} University}
  \country{Sweden}
}

\renewcommand{\shortauthors}{Aler, et al.}

\maketitle

\section{Introduction}
Fairness is a fundamental aspect of justice, and central to a democratic society~\cite{rawls2004theory}. 
It is therefore unsurprising that justice and fairness are at the core of current discussions about the ethics of the development and use of AI systems. 
Given that people often associate fairness with consistency and accuracy, the idea that our decisions as well as the decisions affecting us can become fairer by replacing human judgment with automated, numerical systems, is appealing~\cite{dignum2021myth,abels2021dealing,fernandez2022delegation}. 
Nevertheless, current research and journalistic investigations have identified issues with discrimination, bias and lack of fairness in a variety of AI applications \cite{Mehrabi2021}. 

Algorithmic fairness has been framed as a newly emerging area that studies how to mitigate discrimination in automated decision-making, providing opportunities to improve fairness in AI applications~\cite{dolata2022sociotechnical}.
Research in algorithmic fairness, or AI fairness, has produced a number of frameworks and formal notions of fairness in AI~\cite{Feuerriegel2020,barocas-hardt-narayanan}, many of which are mutually incompatible. 
There is to date no agreement on the relative strengths and weaknesses of such notions, nor on the appropriate scope for their application. Furthermore, as Birhane indicates~\cite{BIRHANE2021100205}: ``many of the `solutions' put forward (1) revolve around technical fixes and (2) do not center individuals and communities that are disproportionately impacted''. 
While technical and formal approaches to fairness remain an active area of research, actual progress is stalled as they insufficiently address the reasons as to why the AI systems were introduced in the first place~\cite{Miceli22}. 
Moreover, they typically fail to take into consideration many of the socio-technical factors that are relevant for a satisfying assessment of the fairness of using AI systems in given situations. 
There is therefore a need to broaden the lens on fairness~\cite{Barabas20,BIRHANE2021100205}, taking the focus away from formal definitions, and allowing for the inclusion of societal and relational aspects to represent how the effects of AI systems affect and are experienced by individuals and social groups~\cite{dignum2022relational}.

In light of these considerations, it is central to try and find tools that tame the complexity of fairness descriptions and models, so as to allow multidisciplinary insights and stakeholder participation that can lead to actionable solutions for the use of algorithms in socially responsible decision-making.
In this paper, we propose such a tool: the ACROCPoLis framework to represent allocation processes with an emphasis on modeling socio-technical and contextual aspects that are relevant to fairness.
For the purposes of this framework, our conception of allocation is very broad, including not only material resources such as food, housing, water, or capital, but also immaterial social constructions such as free time, job posts, legal rights or societal recognition. Consequently, a wide range of situations can be assessed.
The framework provides a shared language in which the factors relevant to fairness assessments for different situations and procedures can be made explicit, including the main entities involved and the relevant interconnections between them.

The goal of this contribution is to redirect the focus to the fact that AI systems are part of wider, complex socio-technical processes where a variety of stakeholders play important roles. 
The stakeholders can for example be social and political forces, as well as technical constraints. 
With a general framework to represent such processes, it becomes possible (1) to compare analogous situations, (2) to highlight the relevant differences in dissimilar situations, and, in cases of conflicting fairness assessments, (3) to capture differing interpretations of the same situation by different stakeholders -- thus organizing the discussion and pointing to the relevant points of disagreement between the parts. As such, our proposed framework for fairness is preparatory to ethical assessments of fairness, describing the stakeholders, their roles, their mutual relations, and the stakes and values at play, while remaining neutral on normative questions.

Although the origin and focus of the here proposed framework are on processes that include AI as part of the decision-making, we see such systems not as independent, neutral technological products disconnected from the context in which they are devised and applied, but rather as artifacts or tools that humans use to shape and enforce social, political, and economic structures. 
The framework is meant to provide an analytical tool that allows a richer appreciation of the complexity involved in fairness assessments by identifying in precise ways the relevant actors, their power to influence the process and the broader context that have all interacted to yield the observed outcomes.
Our overall aim is to initiate an essential discussion on what fairness means within the AI community, which aspects are essential to examine, and which systemic factors one may be overlooking.

\section{Background and Motivation}

 How to best understand fairness is heavily contested in the many areas of study in which the notion is used -- this includes philosophy, political science, social science, and AI and data ethics. 
 Traditional formal operationalizations of fairness stem from fields outside of AI, such as economics. 
 For example, a commonly studied problem in that field is the fair division of goods~\cite{doi:10.1146/annurev-economics-080218-025559}. 
 The questions that these fields of research address, however, are typically dependent on specific formal models, which makes them hard to adapt to a broader societal context. 
 
 Within AI, fairness has been identified as a core principle in a myriad of guidelines and standards focused on the production of responsible and trustworthy AI systems~\cite{EUCommission2018,Chatila-2018}. In this context, fairness is specifically tied to non-discrimination, bias and harm reduction. Thus, the field of AI fairness focuses particularly on the notion of unfair bias. This can be seen in multiple applications: from detecting undesirable word associations in natural language processing (e.g., associating the word ``doctor'' directly with male pronouns) to undesirable associations of features with prediction outcomes in predictive systems (e.g.\ skin color with criminality). Fairness in AI also includes the idea that AI applications should be robust across populations, with similar accuracy and error rates across subgroups (e.g. considering equal representation of people with different socio-economic status in medical predictive systems \cite{xu2022algorithmic}). This can be summarized as avoiding unfair bias in terms of outcomes, benefits and harms. 

Thus, assessing fairness from a technical perspective requires determining how to define and measure undesired bias in terms of specific characteristics, attributes, and outcomes. This is often taken to imply a requirement to quantify all these aspects. Recent years have seen the introduction of several fairness definitions ~\cite{Dwork2012,Hardt2016,Joseph2016,Kearns2019}, capturing different legal, philosophical, and social perspectives. However, there are arguably elements of bias and unfairness that resist quantification, requiring the use of qualitative or mixed methods in order to fully understand the dynamics at play ~\cite{Dignazio2020,Goldfarb-Tarrant2021,Leavy2018,McCradden2020}.

Several statistical operationalizations of fairness, often called ``definitions'', have been developed, defended, and criticized, each hinging on different statistical features of the predictions produced by algorithms as criterial for fairness:  ~\cite{Hedden2021}, for instance, identifies 13 different operationalizations (see also ~\cite{CorbettDavies2018} and ~\cite{Mitchell2021}). Further complicating the issue, analyses have shown that most of such operationalizations are mutually incompatible, and thereby cannot be simply added to improve algorithmic fairness~\cite{Kleinberg2016,Miconi2017}. 
In many cases, trade-offs need to be made between optimizing algorithms for fairness and achieving the social goods which are supposed to be produced by these algorithms~\cite{CorbettDavies2017}.

Such operational approaches to fairness nonetheless constitute a considerable advancement, and provide useful tools for responsible AI designers. However, even setting aside the problems of mutual incompatibility and optimization-performance trade-offs, the fact that they rely purely on quantifiable, statistical measures of fairness is itself problematic. Critical approaches emphasize that fairness is multi-dimensional and that purely quantitative bias definitions and de-biasing methods may lead to new biases, and may be unable to deal with intersectionality (where effects of biases are greater-than-additive, or cause ``double binds''~\cite{Crenshaw1991}). More generally, these definitions tend to ignore or downplay the complexities of social and political contexts, including their complex and mutable dynamics, which makes relying on past data to drive future decisions deeply problematic~\cite{Kearns2019,Fazelpour2020,Fazelpour2021,AlerTubella2022,Blodgett2020,pmlr-v80-liu18c}. Moreover, by seeing algorithmic fairness as a purely technical problem, there is a risk of ignoring the value-laden choices that must be made in the design of algorithms (e.g., what fairness measure to use and what attributes to protect), in how they are applied (e.g., in which social contexts, for what aims, etc.), which individuals and groups should have a say in shaping the algorithms and their applications (e.g., engineers in private companies, democratically elected politicians, affected groups, etc.), and whether algorithmically assisted decision-making is morally acceptable at all in specific cases ~\cite{Wong2020,Selbst2019,Friedler2021,Hull2022,Hoffmann2021,Cifor2019}. 

Following these criticisms, recent proposals have called for more robust frameworks within the field of AI focusing on ``studying up''~\cite{Barabas20,Miceli22,BIRHANE2021100205}, i.e. moving beyond the technical, and including crucial aspects of ``power, historical inequalities and epistemological standpoints''~\cite{Miceli22}. Such undertaking is relevant in the case of fairness, given the diversity of perspectives and models for fairness. The need for a descriptive framework is thereby not a matter of establishing a single agreed understanding of fairness, but rather to provide the means to represent more accurately the elements and relations that should underlie assessments of the fairness of a process and/or situation, and that can help reveal where disagreements lie when conflicting fairness assessments are defended by different parties. 
In this sense, the present work is similar to the framework put forward in \emph{hard choices in AI}~\cite{dobbe2021hard}, which delves into a wider set of socio-technical challenges beyond fairness, but likewise aims to bring to the surface disputes and differing views between different Actors involved in the implementation and use of AI systems.

\section{A Framework for Describing and Analyzing Fairness: ACROCPoLis}
\label{sec:framework}

The goal of our framework is to provide a shared language to specify and parse descriptions of situations and processes that are considered ``fair'' or ``unfair'' (collectively: \emph{fairness statements}). In particular, we  attempt to extract specific features in the description that relate to \emph{why} and \emph{how} the situation or process is perceived as fair or unfair, and to help express this in a clear and consistent fashion for a variety of different situations and processes. 
By means of a common framework, we provide a structured description of the different components of fairness statements, and as such support the understanding of why different Actors may differ in the fairness assessments they provide. 
A common framework also enables a schematic representation of fairness issues in studies on the consequences of AI applications, and enables standardised annotation of fairness-relevant components in meta-studies in AI and ML.

\begin{figure}
    \centering
    \includegraphics[scale=0.61]{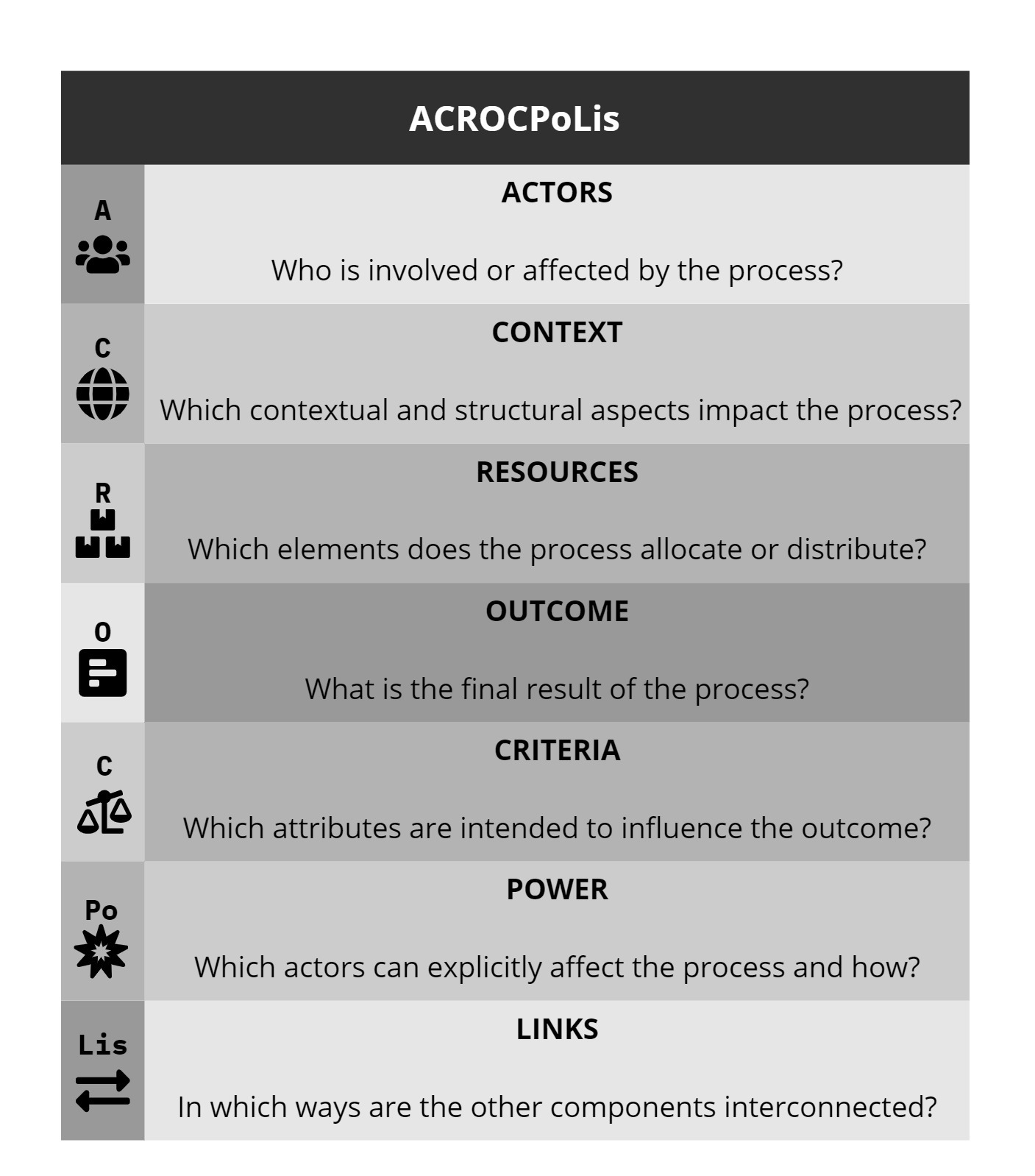}
    \caption{The components of the ACROCPoLis framework, summarized.}
    \label{fig:acrocpolis}
\end{figure}

According to our proposed framework, ACROCPoLis, a process can be decomposed into seven components: %
\textbf{A}ctors,
\textbf{C}ontext,
\textbf{R}esources,
\textbf{O}utcome, 
\textbf{C}riteria,
\textbf{Po}wer,  and  
\textbf{Li}nk\textbf{s}, depicted in Figure~\ref{fig:acrocpolis}. Briefly, \emph{Actors} are the agents of the situation/process under study, whether they are individuals, groups of people, institutions or a combination thereof.
\emph{Context} captures the relevant contextual and structural factors that bear on the specific situation/process, as well as information about what the situation or process involves. 
Fairness processes and situations often involve reasoning about the (re)distribution of specific \emph{Resources}, which we take in a broad sense to include not just material items such as money and food, but also increased representation, compensation, agency, legal rights, acknowledgment, etc. 
The redistribution of such Resources are one type of \emph{Outcome}. 
Another Outcome type is decision, for instance regarding access to social benefits, hiring, and in legal contexts.
\emph{Criteria} contains whatever attributes are used to influence the Outcome, and \emph{Power} describes the ability of each Actor to influence the system, including the nature of the influence.
The \emph{Links} highlight important connections between categories and concepts in the described system, in particular those connections that are not obvious from the previous description.

The choice of categories responds to the growing literature calling for a multi-disciplinary and socio-technical view of processes that involve AI systems. 
Considering a broad category of Actors allows for including companies and organizations, which often hold accountability for the Outcomes of the process~\cite{Raji20}, as well as communities and individuals affected by a given process, whose perspectives are key to fairness assessments~\cite{BIRHANE2021100205}. Following this human-centric perspective, it is crucial to include aspects of the Context to allow for discussion of contextual and structural factors that are often the root cause of unfair processes, through perpetuation or exacerbation of existing inequalities~\cite{buolamwini2018gender,kiritchenko2018examining}.

In terms of algorithmic fairness, the focus is often placed on allocation processes~\cite{Verma2018}, where Resources are being distributed. We consider Resources in a wide sense, including ``labels'',``scores'' or even ``error rates'', but also aspects such as recognition or representation. 
This allows for modeling part of the Outcomes of a decision process as a reallocation of Resources, for which different definitions of fairness can apply. As our focus is on cases in which AI systems are integrated in socio-technical processes, a natural category to include is the Outcome of the process being considered. The effects of Outcomes on the Actors, the Context and Resource allocation play a key part in assessing the (un)fairness of a process. 
When describing a process in terms of fairness, a focus is often on the attributes of the Actors and the Context that explicitly influence the process~\cite{Verma2018}. Identifying these aspects, which we name Criteria, allows for an explicit characterization of what concretely produces the Outcomes observed.

A final critical choice in the framework we propose is to explicitly include a category on Power. 
This category aims to make explicit which agents have direct influence, i.e. Power, over the process, as well as to describe how much agency (or lack thereof) each Actor possesses. 
Being explicit on this aspect is critical for a sociological view of processes~\cite{Barabas20,Miceli22}, since it is crucially intertwined with issues of fairness (who holds the power?~\cite{BIRHANE2021100205}) and accountability.

Describing the above categories is, in many cases, not sufficient to assess or describe the fairness of a process. 
Indeed, fairness assessments often hinge on the relationships between components. For this reason, Links, denoting such components, are an integral part of the framework.

The next sections detail the different elements of ACROCPoLis. As a running example we use the well-known COMPAS case, where an AI decision support system ostensibly evaluated the recidivism risk of court defendants based on various demographic data. In an exposé by ProPublica, the system was found to yield different results based on race, in effect recommending harsher penalties and higher bail payments for Black people than white people~\cite{Mehrabi2021}. In particular, the perspectives of both Northpointe (the company that built the COMPAS system) and ProPublica will be described, as well as the factors these Actors took as most relevant, revealing the differences in the fairness assessments they provided.

\subsection{The Components}

\paragraph{Actors} When describing a process as being (un)fair, there is always a set of agents that make and/or are subjected to certain decisions and/or allocations of Resources. In particular, such a description considers these agents as \emph{moral agents}, with their circumstances and feelings being worthy of consideration in assessing the fairness of a situation, and their decisions and actions having moral weight. As such, current AI agents are unlikely to be considered Actors under this framework, as current automated systems are generally not considered moral agents. However, it is not uncommon for groups or institutions to be considered moral agents in and of themselves, and as such, they should be represented as Actors in this framework. Furthermore, Actors are not simply identifiers but are differentiated by having certain attributes, which can be used to study and describe a large number of Actors' involvement in a process by way of grouping them by their various attributes.

For our running example, both Northpointe and ProPublica consider the relevant Actors in the system to be Northpointe themselves as designers of the COMPAS system, the policymakers who approved its use, the judges who used the system to inform their decisions, and the defendants who were subjected to it.

\paragraph{Context} Although it is possible to talk abstractly about various conceptions of fairness, most fairness statements concern some specific (type of) process embedded in some Context. In particular, we take this category to contain both the specific context that the fairness statement concerns, and any of its aspects connected to the (re)distribution of Power and Resources, including structural and population-level aspects. The concept Context also includes the sociotechnical systems active in the process, relevant aspects of the dynamics of such systems, as well as other aspects of the world that are relevant for making assessments about the (un)fairness of the process.

In relation to the COMPAS system, ProPublica and Northpointe have different views on important parts of what Context to include in the description of the system. In particular, while existing inequalities are acknowledged by both, the impact of racism on the higher arrest and conviction rates of Black people in the training data is particularly important to ProPublica's description of the situation. Conversely, an important part of the Context for Northpointe is that the limited Resources available to the court motivate using the COMPAS system as a time-saving measure in order to process more court cases faster.

\paragraph{Resources}
Our conception of Resources is very broad, including not only material resources such as food, housing, water, or capital, but also immaterial social constructions, such as dignity, free time, and legal rights. 
Moreover, we also include concepts such as societal recognition and agency within a structure as Resources. 
This category therefore includes any element that can be seen as being (re)distributed by the process. 
In most cases, the description of a process as fair or unfair relates to some distribution of Resources in this broad sense, either with the Resource distribution impacting the Outcome of the process, or that the process, through redistribution, leads to, counteracts, or perpetuates some existing (un)fair distribution of Resources.

The Resources relevant to the description, for both Northpointe and ProPublica are Resources taken from, on the one hand, incarcerated people like ``bail'', ``time not in prison'', and social status in general.  On the other hand, Resources  are also taken from the state or society at large like ``state funds'' as applied to jails, prisons, courts, and the costs of social measures for recidivism prevention. In this wider view, the issues of private prisons, prison labor and prisoner exploitation also become relevant, not only due to higher (absolute) incarceration rates leading to a redistribution of Resources from the public into private hands, but also due to the fact that uneven incarceration rates between groups can lead to exploitation of the incarcerated group, redistributing the fruits of their labor and, quite often, their savings and future earnings (through debt), to beneficiaries in other groups.

\paragraph{Outcome} When describing fairness in processes, the Outcome is clearly relevant. Outcomes can be modeled as a (re)distribution of some Resource, a (re)definition of a Power relation, or the introduction or removal of Power or agency invested in some Actor. It is also possible to model Outcomes on the group level, and across different timeframes; a process run once, or in a perfect society, may be perfectly fair, but under existing societal structures and run over many iterations, it may entrench or introduce unfair (dis)advantages or distribute Resources unfairly. This category therefore includes changes, effects or results that are direct consequences of the process being analyzed.

Both Northpointe and ProPublica have an individualist view on the Outcomes of the system, focusing on each bail judgment and prison term for individual defendants. However, both also look at the process from a statistical, iterated perspective, comparing the Outcomes over many individual runs. In particular, a critical Outcome is the \emph{rate of false positives} in the group of Black defendants, which is higher than the same statistic for white defendants.  To clarify the notion of Resource distribution with this example, COMPAS redistributes among the US population Resources such as freedom, political participation, perceived social value, economic opportunities, and the like, in light of the sentencing decisions it contributes to.  

\paragraph{Criteria}  When describing a process in terms of fairness, a focus is often on the attributes of the Actors and the Context that explicitly influences the process. 
These denote the ``causality'' behind the system, insofar as they capture the criteria underpinning the decisions that produce Outcomes. As such, the criteria form a connection from the state of the world (as defined by the Actors, their attributes, and the Context the process is enacted within) to the Outcome of the process. Criteria may be explicit or implicit, intended by the Actors in exercising their Power, or unintended.

In the case of COMPAS, the Criteria behind the process consist of i) the various types of demographic and judicial data that COMPAS employs in its algorithm; ii) the sentencing guidelines used by judges; iii) the instructions about how judges are to use COMPAS; and iv) the actual use of such guidelines and instructions. These Criteria are agreed upon by both ProPublica and Northpointe. It is notable that in both views, the race of defendants should not be part of the Criteria.

\paragraph{Power}
Power describes the explicit ability of an Actor to affect the process being described. This category aims to make concrete which agents have a direct influence, i.e. power, over the process, as well as to describe how much agency (or lack thereof) each Actor possesses. In particular, an Actor can have Power over various aspects of the Context in which the Actors exist, as well as over the Criteria by which a fairness process is conducted or evaluated. Additionally, Actors may have Power over other Actors, and thus second-order Power over any aspects that those Actors control.

As noted in the Actors section, various Actors in the system were able to control specific parts, though generally Northpointe was keen on downplaying the amount of control they could enact through their design of the COMPAS system, and ProPublica moreover was clear that public officials had the Power to limit the type of information that the system could be provided with. Both parties agreed that the judge was the one who ultimately had the Power to use or ignore the COMPAS system in making their judgments. For ProPublica, however, COMPAS retains an undue degree of Power in influencing judges' decisions, who may be more prone to follow the outputs of the system, despite instructions not to give too much weight to them in informing their decisions.

\subsection{The Links}

\begin{figure}[ht]
    \centering
    \includegraphics[width=0.8\textwidth]{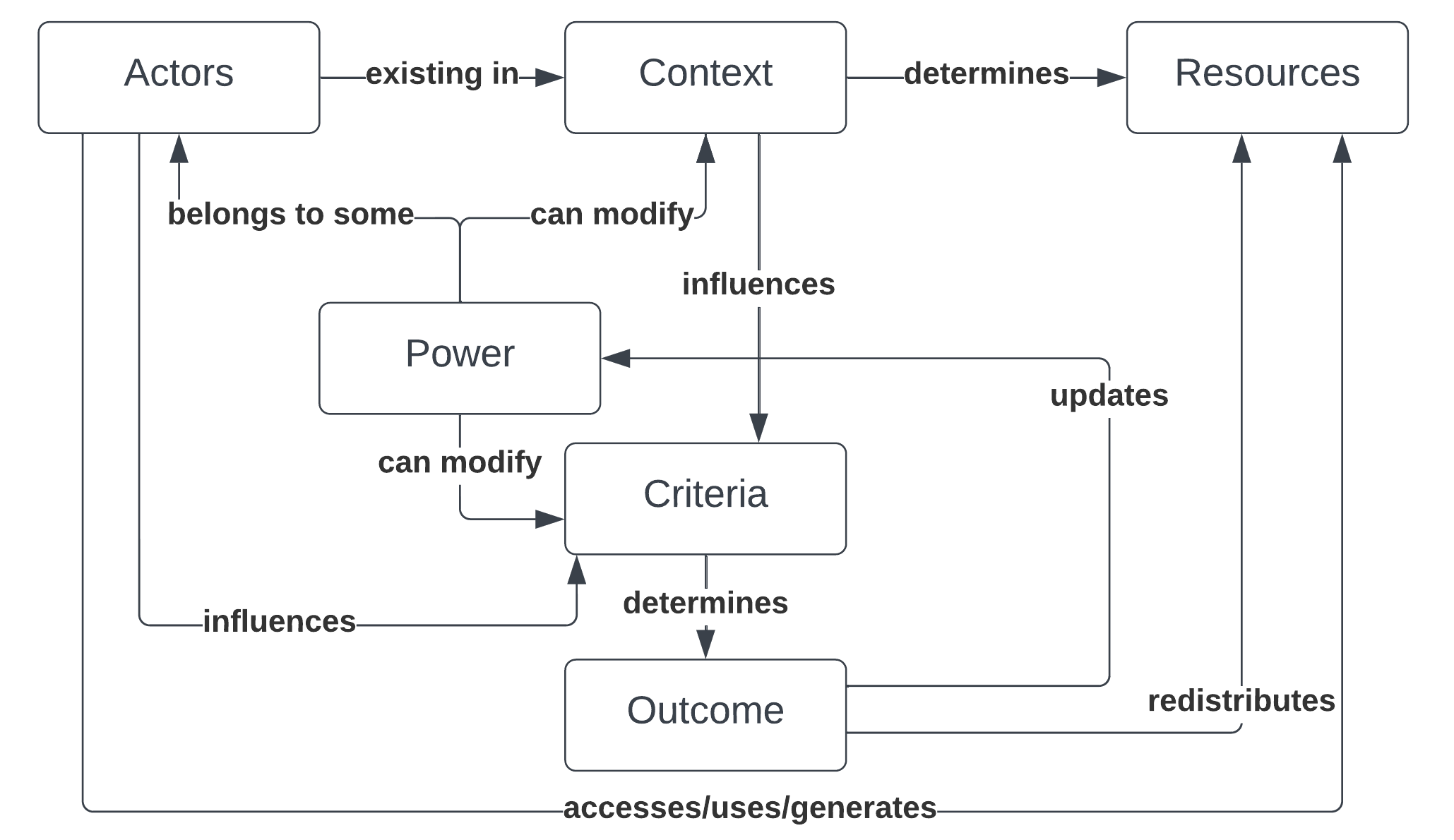}
    \caption{A schematic overview of the components and links of the ACROCPoLis framework}
    \label{fig:acrocpolis_chart}
\end{figure}

Describing the various interconnections between components of the system is a key part of any application of ACROCPoLis. 
The Links can be understood as shown in Figure \ref{fig:acrocpolis_chart}: Some certain Power belongs to the Actors that can use it, and the Power can modify the Context that the Actors exist in.
The Context and the Actors both influence the Criteria while the Power modifies the Criteria -- the Criteria in their turn determine the Outcome. 
An Outcome can update the Power, and also redistribute the Resources that are determined by the Context and used, accessed or generated by the Actors. 

While the Links indicated in Figure \ref{fig:acrocpolis_chart} are intended to serve as a good basis for discussion and analysis, they may need to be further expanded to fully describe the connections an analyst considers to be relevant to assess the fairness of the process under study. For example, Actors may have Power over other Actors, either directly or through the existence of hierarchical power structures in the Context, which would give these Actors second-order access to the Powers that their subordinates have, or there may be more indirect connections between Actors, Context and Outcome than is appropriate to model as a criterion, but that is nevertheless relevant for assessing the fairness of the process. 
These more fine-grained Links may be added to the ACROCPoLis analysis on a case-by-case basis.

As mentioned under Outcome, an important additional connection in the use of the COMPAS system is that between the race of the defendant and the COMPAS assessment (in particular the rate of false positives), which also often influences the final decision by the judge. In their descriptions of the situation, both Northpointe and ProPublica agree that such a difference existed, though they differ on whether it is relevant or not.

\subsection{The Evaluation} 
The above framework is, to a certain extent, purely descriptive. It picks out relevant aspects of processes and situations under analysis, highlighting those components that are central to the assessment of the process as being fair or not. A crucial part of any fairness analysis is, however, normative: one needs to decide whether the situation, as presented, is fair or not, and why. What Links are missing? What aspects of Context and Power are ignored? Are there conflicts between what the Criteria actually are and what they were intended to, or should, be? 

Such an evaluation can be obtained by comparing ACROPoLis analyses based on the perspectives of stakeholders with diverse and potentially diverging viewpoints. 
Ideally, the ambition is to present an integrated analysis that is as complete as possible. 
In this way, conflicts and disagreements about the nature of the different components of the process or situation can be brought to surface, facilitating a clearer assessment of the case and of the contested points, opening the way for compromise and/or consensus building. 
The final result can then be accompanied by a joint concluding assessment with respect to the fairness of the outcome and, in case actions to address fairness issues are supposed to follow, by a counterfactual ``future state'' ACROPoLis proposal that suggests how these issues can potentially be addressed. 

These features of ACROCPoLis can be clearly seen in the dispute between Northpointe and ProPublica. In defending the fairness of the system, Northpointe pointed to the existing differences in recidivism rate between Black and white defendants as an explanation for why the rate of false positives are expected to be higher for Black people than for white people, purely on the basis of statistical and demographic considerations. COMPAS, therefore, would be correctly mirroring historical and demographic patterns, which it was not intended to change. To this, ProPublica objected that COMPAS perpetuates patterns of unequal treatment of the Black population by the justice system, seeing COMPAS as partly responsible not only for failing to ameliorate, but also for worsening such unequal Outcome. 

In other words, the disagreement between Northpointe and ProPublica is largely due to the different views each party takes both to the Power COMPAS exercises, the kinds of Criteria that such a system should employ, and thus the Outcomes it should produce.

\begin{figure}[ht]
    \centering
    \includegraphics[width=0.8\textwidth]{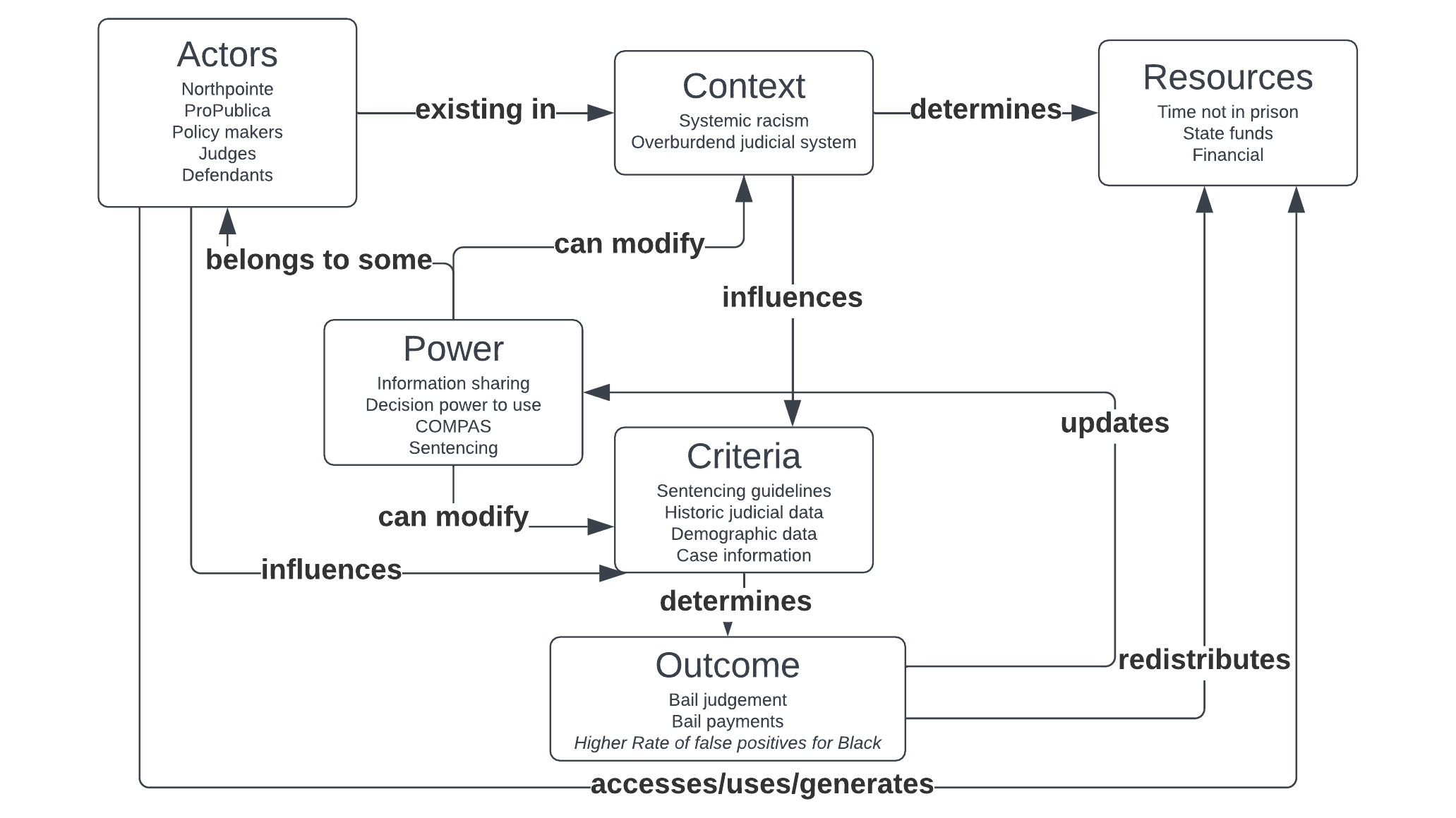}
    \caption{The COMPAS case study visualized using our framework, based on the components identified in~\cite{Mehrabi2021}. We can use the Links between them to identify where the process lacks fairness.}
    \label{fig:compas:diagram}
\end{figure}

Figure~\ref{fig:acrocpolis_chart} summarises, with a degree of simplification, the COMPAS case study examined above, following the framework provided by ACROCPoLis.
Figure~\ref{fig:compas:diagram} shows the resulting instantiation.

\subsection{Using the framework}

Beside describing in precise ways the fairness-relevant elements and relationships in situations and processes, ACROCPoLis can also be used to identify the crucial factors leading to unfair outcomes, and the most appropriate loci for intervention. For instance, in the COMPAS example, it becomes clear that the Context of systemic racism against Black people in the US strongly influences the Outcomes, despite the fact that the Criteria were supposed to be neutral toward defendants' race. Thus, an intervention to ameliorate the situation could involve modifying the Criteria to counter, instead of mirroring, such systemic and historical inequality of treatment between groups in the US population. 

Analogously, defendants have diminished autonomy and Power insofar as judges' decisions are influenced by the groups they are seen to belong to, risking to downplay the individual history and circumstances of defendants in motivating the decisions. Such considerations put pressure on any system that uses historical demographic data to help decide the fate of individuals, suggesting that a larger societal discussion may be needed when evaluating whether AI technologies such as COMPAS should be used by governments.

\section{YouTube Recommendation System: A Case Study}
To further illustrate the use of ACROCPoLis, let us examine recent work on the search and recommendation algorithms of the video platform YouTube, and how they are analyzed in terms of fairness in the literature.
YouTube's algorithms are proprietary (closed-sourced and not documented in a transparent manner from a user or public perspective). 
This opacity has triggered studies into the features of the algorithms as well as the consequences of their operation within the platform, based on publicly accessible facts~\cite{arthurs2018researching,Stray-2022}.

Two such studies are examined here: the first study focuses on examining the question of whether the YouTube recommendation algorithm favors a small subset of the video content available on the platform ~\cite{10.1007/978-3-030-78818-6_10}; and the second examines whether YouTube's search algorithm is biased toward video content associated with specific positions on the United States political spectrum~\cite{9660012}.

Kirdemir et al.~\cite{10.1007/978-3-030-78818-6_10} investigate the structure of recommendation networks, and probabilistic distribution of video recommendations from a given node in the network. 
Recommendation graphs are constructed based on eight different real-world scenarios (seed video sources) to then apply a stochastic approach and observe PageRank~\cite{Page-1998,Page-1999} distributions over such graphs. 
The study found that a small fraction of the nodes in the recommendation network received the large majority of connections, suggesting that YouTube's recommendation algorithm is biased toward a small number of content items available on the platform.

Lutz et al.~\cite{9660012} examine political bias in YouTube's recommendation and search algorithms from a US-centered, binary (right-leaning vs left-leaning) perspective. 
Search and recommendations were studied in separate experiments. We will be concerned exclusively with the study on the search algorithm. For that experiment, the authors scraped the 200 top search results for a variety of politically-charged terms in the US public debate, using four YouTube accounts with different viewing profiles create for the experiment. The top search results for each term and each profile were evaluated for their political leaning. The study found that left-leaning content items were significantly more likely to appear among the three top search results.

Do these studies indicate unfair biases in YouTube's search and recommendation algorithms? Let us use the ACROCPoLis framework to examine these cases in detail, starting from the identification of the relevant components.

\begin{table*}[htb]
    \small
    \centering
    \caption{ACROCPoLis analysis of the YouTube recommendation and search algorithms, based on results from two different papers, one focused on content bias, one on political bias. Items between parentheses are neglected in the papers, but need to be included in a full ACROCPoLis analysis
    }
    \label{tab:youtubeexample}
\begin{tabular}{|l|p{0.45\textwidth}|p{0.45\textwidth}|}
\hline
   & \textbf{Content Bias~\cite{10.1007/978-3-030-78818-6_10}} 
   & \textbf{Political Bias~\cite{9660012}} \\
\hline
A 
& Content creators, content consumers, YouTube, (policymakers) 
& Content creators, content consumers, YouTube, (policymakers) \\
\hline
C 
&
Competition between content creators for user views; abundance of content items; YouTube's revenue model 
& Population of platform users with diverse positions in the political spectrum; social and legal protections of free speech
\\
\hline
R 
& Time spent on the platform by users; views per content item 
& Time spent on the platform by users; views per content item
\\
\hline
O 
& Users following platform recommendations lead to the same small group of content items; 'winner takes all' effect 
& Users more exposed to left-leaning content items in top search results in the US \\
\hline
C 
& Proprietary recommender algorithm, criteria unknown
& Proprietary search ranking algorithm, criteria unknown \\
\hline
Po 
& Content creators: produce content and add it to the platform  \newline 
(Content consumers: choose what content items to view and engage with, and for how long) \newline
YouTube: generate content recommendations \newline
& Content creators: produce political content and add it to the platform \newline
(Content consumers: choose what content items to view and engage with, and for how long) \newline
YouTube: generates ranked search results \\
\hline
Lis 
& See remainder of this section.
& \\
\hline
\end{tabular}

\end{table*}

In Table~\ref{tab:youtubeexample}, the ACROCPoLis framework is used to describe the fairness-relevant components extracted from both articles.
In addition to identifying the components playing a role in each paper, let us take a closer look at the \textbf{Links} that are relevant for assessing the fairness of the situations examined in the studies at hand, and what they reveal about the limitations of the analyses proposed therein.

In~\cite{10.1007/978-3-030-78818-6_10}, the focus is on the possibility that the YouTube recommendation algorithm is biased toward certain sorts of content, thus generating a pattern in which users are ultimately pushed towards a small fraction of the content available on the platform (Content Bias). 
The Actors identified by the paper, i.e., content creators, content consumers, and YouTube itself, are seen on the background of YouTube's revenue model (Context). 
The platform revenue's model is largely organized around advertisement revenue, which is shared between the company and the content creators. 
Thereby, YouTube have an interest in maximizing the time spent by consumers on the platform, thus increasing the amount of advertisement they are exposed to; while content creators have an interest in maximizing the number of platform users that view the content items they produce, thus increasing the share of advertisement revenue they receive from the platform (Resources).

Interestingly, the paper seems to place most of the Power in the hands of YouTube itself, insofar as the platform is responsible for the recommendation algorithm that increases the visibility and accessibility of certain content items rather than others. 
The ACROCPoLis framework, however, invites an examination as well of the Power, if any, that the other Actors possess. First, content creators arguably have an important amount of Power in shaping the allocation and distribution of the relevant Resources (time spent and views by content consumers). Indeed, it is at least in part the attractiveness and interest of the content items they produce that attract consumers to YouTube to start with. Moreover, content creators may aim at creating content items that they believe will attract a larger number of users, in light of their own knowledge about societal trends and public interest in different subject matters. Finally, content creators may partially reverse-engineer some of the Criteria used by YouTube's recommendation algorithm in order to increase the likelihood of appearing as recommended content to the population of consumers they think might be more attracted to their content items (a sort of Recommender Engine Optimization). In consequence, content creators should arguably not be seen as passive, powerless Actors in the situation under examination. While their interests partially overlap with those of the platform, they also partially differ, insofar as content creators compete among themselves for the available Resources, while YouTube is mostly interested in overall content consumption on the platform.

Similarly, and perhaps more strikingly, content consumers in the platform are treated as purely passive Actors in the paper. Indeed, the methods used in the research simulate consumers that fully follow the deliverances of the recommender algorithm (with some noise added), thus being largely deprived of any Power in deciding what content to consume, and for how long. This is an oversimplification that the authors acknowledge, but it risks neglecting relevant Links that are central to the assessment of the fairness of the outputs of YouTube's recommender algorithm. 
Content consumers have at least some degree of autonomy in selecting what content items to consume, and which recommendations to follow or to ignore. Importantly, they may thus influence the Criteria embodied in the recommender algorithm itself. 
It is indeed in the interest of YouTube to generate recommendations that mirror the type of content that consumers may want to engage with, and it is hence to be expected that the algorithm responds to the patterns of consumer preference in the platform. 
Moreover, the large amount of data that Alphabet, the owner of YouTube, possesses about each user of its services allows considerable personalization of content recommendations. 
However, as the authors admit, studying the influence of personalization on the recommendations produced by YouTube is very challenging. 
In brief, through the Power content consumers possess by means of their consumption preferences, they likely influence the Criteria (i.e., the features used by the recommendation algorithm) that lead to the observed Outcomes (i.e., recommendations that encourage consumers to view a small fraction of the content items available). 

The above considerations suggest that an assessment of whether or not the content bias found by the researchers is unfair or else requires a fuller picture of the Power the relevant Actors possess, and how they shape the Criteria that produce the content bias found by the study. 
It may well be, for example, that no unfairness is involved, say, if what the recommendation algorithm does is to successfully draw users toward content items that they are more likely to want to consume. 
Given the varying quality of the content items available, the variety of subject matters treated in such content items, and social trends that help determine what is or is not in the 'public eye', it is arguably expected that relatively few content items are taken to be particularly interesting by content consumers. 
In other words, YouTube's recommender algorithm might be appropriately promoting content items with higher relevance, interest, and quality. 
Importantly, we are not claiming that this is the case. 
Our claim is merely that the paper leaves underdefined several ACROCPoLis components and Links that are crucial to a satisfying assessment of the fairness of the situation being examined.

Analogous considerations apply to~\cite{9660012}. 
The result of the paper that concerns us here is the finding that YouTube's search algorithm shows a bias toward placing more left-leaning content among the top three search results on the platform than would be expected in light of the relative representation of political leanings among content items.
One central methodological choice in the paper is that of inferring the distribution of political leanings across the YouTube user base in the US from the political leanings found in the sample of content items examined in the study. 
Under the ACROCPoLis framework, this methodological choice is far from inconsequential, since it involves a conflation between two different Actor categories that are non-overlapping and that have partially incompatible interests and Power, namely content creators, and content consumers. 
Content creators are plausibly interested in maximizing the viewership of their content items, and content consumers are interested in consuming the content items that they find most interesting, entertaining, or informative. 
While this suggests that content creators should aim at satisfying consumer preferences, the goals of content creators can also involve other considerations, such as popularizing minority positions, spreading relatively neglected ideas and ideologies, and the like. 
In other words, it is plausible that the political leanings expressed by content producers are not an appropriate measure of the political leanings of content consumers. 

As the ACROCPoLis analysis makes clear, content creators and content consumers are different Actors, who possess different kinds and degrees of Power in shaping the search ranking algorithm (Criteria) and thus the Outcomes. 
Thereby, the study results could at most be used to argue that there is an unfair bias in top 3 search results in the US under a uniform distribution understanding of fairness (i.e., equal exposure to be given to each political leaning, regardless of representativeness in the population). 
It is doubtful, however, that uniform distribution is the appropriate approach to fairness to be used in this case, as it would also require YouTube to give space to extreme political positions that have little representation in the general population.

As with~\cite{10.1007/978-3-030-78818-6_10}, this study does not take into consideration the role of content consumers in shaping the search ranking algorithm that YouTube employs. 
As the authors admit, this is a limitation of the study, for if the political bias found is mostly due to consumer preferences, it is debatable that such a bias is unfair at all, rather than being an expression of the Power exercised by content consumers over the search ranking algorithm (and thus over YouTube itself) by means of their autonomous consumption choices. 
Both papers, moreover, do not go into the role that policymakers have in shaping the Criteria, and thus the Outcomes. This neglect is arguably justified, insofar as, at least in solidly democratic countries, policy interventions are mostly concerned with illegal and harmful content, while free speech guarantees make it so that policymakers have little Power when it comes to influencing the Criteria and Outcomes involving legal content items. 

Importantly, the considerations above are not meant to diminish the value of the studies examined. 
They are merely intended as illustrations of how the ACROCPoLis framework can help furnish a fuller picture of what pieces of information are needed in order to provide more strongly substantiated fairness assessments of specific situations and processes. 
Indeed, the ACROCPoLis analyses above point out limitations in the studies examined, thus revealing further research questions that need to be explored to complement their findings. 

Finally, it should not have escaped the reader's attention that a glaring gap in the foregoing examination, and in the studies themselves, is the Criteria component. 
This is due to the fact that YouTube's algorithms are proprietary and closed, making it so that the Power that different Actors exercise in shaping it, as well as how the algorithms lead to the observed outcomes, is only partially inferrable by input-output testing of the platform. 
Moreover, it is likely that YouTube's algorithms are constantly in flux, with fixes and tweaks introduced by the company to improve their working in light of the platform's interests and the regulatory requirements it is called by policymakers to respect. 
This makes it so that a fully adequate fairness assessment of YouTube, by the lights of ACROCPoLis, cannot currently be produced. 
As an aside, it is worthwhile to point out that the counterfactual scenario in which YouTube's algorithms are made openly available, either by the company or by successful reverse-engineering, would importantly change the ACROCPoLis analyses. 
After all, it is plausible that in such a situation more of the Power would move to the hands of content creators, hence changing the Outcomes produced (potentially in undesirable and/or unfair ways).

\section{Concluding Remarks}

With ACROCPoLis, we have proposed a common, uniform model to represent and analyze fairness statements. 
In this paper, we highlight the potential for practical use with an analysis of the well-known COMPAS case, as well as analyses of two recent studies focused on biases in YouTube. 
Still, ACROCPoLis is merely a starting point on the road toward operationalizing algorithmic fairness modeling. 
Further applications are needed to assess, validate and improve ACROCPoLis in order to resolve potential weaknesses, and to formulate extensions as well as refinements to better catch conceptual nuances that may be of substantial practical relevance. 

ACROCPoLis is mostly a descriptive framework that supports the identification of the relevant aspects to take into consideration in fairness assessments.
A crucial part of any fairness analysis is, however, normative: one needs to decide whether the situation, as presented, is fair or not, and why. 
In particular, it is necessary to identify which fairness questions or situations do not fit into the framework as it stands. Of particular importance is the identification of aspects that are relevant for the specification of fairness assessments that ACROCPoLis may be leaving out.
Our next step in this direction is to apply ACROCPoLis to the domain of law and legal reasoning~\cite{surden2020ethics}, which has a long tradition of developing frameworks and theories that facilitate fair and ``ethical'' decision-making, and whose application is central to fundamental societal challenges such as the facilitation of democracy and human rights.

Further work is needed to address the following issues (this list is non-exhaustive, but intended to serve as a starting point for future research and dialog across disciplines and stakeholders):
\begin{itemize}
    \item
    Is the framework flexible enough to maintain its adequacy as AI-based technology and associated fairness concerns evolve?
    \item To what extent does ACROCPoLis support interdisciplinary communication and public discourse, considering that conceptions of fairness and how fairness should be modeled differ substantially across fields and communities?
    \item What are the challenges around operationalizing the framework at scale in real-world socio-technical systems?
    \item Does ACROCPoLis support procedural notions of fairness~\cite{https://doi.org/10.1111/j.1744-540X.2007.00458.x}, in which the fairness of the \emph{process} is put into focus, i.e.,\ where procedural aspects of a decision matter just as much or more than the consideration of its inputs and outputs?
    \item How can ACROCPoLis accommodate the potential future scenario of AI systems that attain the status of moral agents?
    \item  Is a framework such as ACROCPoLis sufficient, and useful, to handle questions ``beyond'' fairness, in ways that support cross-disciplinary communication?
\end{itemize}

Addressing these issues will require a cross-disciplinary and participatory approach, in which the aim of a shared understanding of how to interpret concrete fairness situations is central.  
Moreover, we must point to the central role of modelers in the fairness debate. 
Indeed, model creators have the power to frame how fairness is assessed, and thus the outcome of fairness evaluations. 
For example, Power (or proxies thereof, such as money) may be considered a Resource (which implies it can be distributed), or it may be modeled as an Actor's attribute, which may obfuscate the possibility of redistributing power as a potential solution to an unfair situation. Importantly, a risk that needs to be further examined is whether frameworks like ACROCPoLis may lead to the neglect of relevant perspectives that are distant from the ones of the modelers themselves. 

Finally, we contend that the conceptual assessment and refinement of ACROCPoLis is in fact more important than its mathematical formalization or its implementation in an IT system-executable format: societal validation must precede technical verification, or the creation of a verifiable specification. 
Moreover, there is a growing debate about whether fairness is sufficient as a criterion when it comes to addressing algorithmic harms, with scholars calling for an orientation towards justice and dismantling oppressive structures~\cite{Cifor2019,Bennett2019,Hoffmann2021}.

A key idea for future work is to apply ACROCPoLis at the design stage of intelligent systems. 
By modeling what role the system will play within a wider process, the clarity of the framework allows us to pose key questions about its context, purpose and effects, including whether an AI system is needed in the first place. These considerations have been dubbed ``hard choices'' \cite{dobbe2021hard}, pointing to an ``overlap between design decisions and major sociotechnical challenges'', which frameworks such as ACROCPoLis can help clarify.
\newpage

\bibliographystyle{plain}
\bibliography{bib}

\end{document}